\documentclass[prd, twocolumn, nofootinbib]{revtex4}

\usepackage{epsfig} 
\newcommand{\beq}{\begin{equation}}
\newcommand{\eeq}{\end{equation}}
\newcommand{\beqa}{\begin{eqnarray}}
\newcommand{\eeqa}{\end{eqnarray}}
\newcommand{\om}{\Omega_m}
\newcommand{\swp}{\sigma_{\_}(w')}

\def\la{\mathrel{\mathpalette\fun <}}
\def\ga{\mathrel{\mathpalette\fun >}}
\def\fun#1#2{\lower3.6pt\vbox{\baselineskip0pt\lineskip.9pt
  \ialign{$\mathsurround=0pt#1\hfil##\hfil$\crcr#2\crcr\sim\crcr}}}

\begin{document} 

\title{Is Dark Energy Dynamical? Prospects for an Answer} 
\author{Eric V. Linder \& Ramon Miquel} 
\affiliation{Physics Division, Lawrence Berkeley Laboratory,
Berkeley, CA 94720} 
%\email{evlinder,rmiquel@lbl.gov}

\begin{abstract} 
Recent data advances offer the exciting prospect of a first 
look at whether 
dark energy has a dynamical equation of state or not.  While formally 
theories exist with a constant equation of state, they are nongeneric -- 
Einstein's cosmological constant is a notable exception. 
So limits on 
the time variation, $w'$, directly tell us crucial physics.  
Two recent improvements in supernova data from the Hubble Space 
Telescope allow important steps forward in constraining the dynamics of 
dark energy, possessing the ability to exclude models with $w'\ga 1$, 
if the universe truly has a cosmological constant.  
These data bring us much closer to the ``systematics'' era, where 
further improvements 
will come predominantly from more accurate, not merely more, observations.  We 
examine the possible gains and point out the complementary 
roles of space and ground based observations in the near future.  To 
achieve the leap to precision understanding of dark energy in the next 
generation will require specially designed space based measurements; 
we estimate the confidence level of detection of dynamics (e.g.\ 
distinguishing between $w'=0$ and $w'=1$) will be 
$\sim1.8\sigma$ after the ongoing generation, improving to more 
than $6.5\sigma$ in the dedicated space generation. 

\end{abstract} 

%\keywords{dark energy --- cosmology:observations --- 
%cosmology: theory --- supernovae}

\maketitle 

%\date{\today}

\section{Introduction} \label{sec.intro}

The acceleration of the expansion of the universe represents a critical 
mystery.  To the physicist this raises questions about Einstein's 
general relativity, quantum physics, and extra dimensions.  To the 
interested spectator it contains the determination of the fate of the 
universe. 

We will not learn a unique answer to all these questions without an overall 
theory for the dark energy causing the acceleration -- one possibility 
for which is Einstein's cosmological constant.  But a major question 
that is directly accessible is whether the dark energy is dynamical, 
possessing a variation in time, sufficient in itself to rule out a 
cosmological constant.  Observations of 
the expansion, most clearly and directly at present through the use 
of Type Ia supernovae as calibrated candles for the 
distance-redshift behavior, can constrain the physics through 
the measured properties of the dark energy. 

These properties are conveniently described by the equation of state 
ratio of dark energy pressure to density, determining how the dark energy 
evolves with cosmic time $t$ or redshift $z$, and written $w(z)$.  Generically 
there is such an evolution, though the cosmological constant possesses an 
unchanging $w=-1$.  It is difficult to estimate a typical value for the 
rate of change without a fundamental theory.  But let's make what appears 
to be a reasonable assumption that the dark energy behavior responds to 
the cosmic expansion $a(t)$, so the equation of state changes 
on a timescale related to the transition from matter domination 
to dark energy domination, or slower due to the influence of 
the remaining matter.  Then a variation less 
than or of the same order of the 
Hubble expansion rate $H=\dot a/a$ seems natural: 
\beqa
\dot w &\equiv& \frac{dw}{dt}=H\frac{dw}{d\ln a}\lesssim H, \\ 
w' &\equiv& \frac{dw}{d\ln a}\lesssim 1. \label{eq.wdot}
\eeqa 
We caution that our notions of naturalness do not constrain the universe 
to actually act that way! 

In \S\ref{sec.data} we examine what the quality of the most recent 
supernovae data can 
tell us about the magnitude of $w'$.  We emphasize that this is an 
illustrative look, based on estimated {\it types} of data numbers and 
accuracy, rather than the actual data themselves.  Possible 
future samples involving more data and better data are examined in 
\S\ref{sec.future} to see how they give the greatest leverage on 
improving our knowledge of 
the dynamic nature of dark energy.  We conclude in \S\ref{sec.concl} 
with thoughts 
on what new advances are required to understand the fate of our universe. 

\section{Data and Dynamics} \label{sec.data}

For constraining the nature of dark energy, we consider three forms 
of data currently available.  Type Ia supernovae (SN) magnitudes and redshifts 
play a central role in tracing the expansion history of the universe, 
and led to the discovery of the acceleration \cite{perl99,riess98}. 
When expanding the description of dark energy to include not only the 
contribution of its energy density but its present equation of state 
$w_0$ and a measure of its time variation $w'$, other 
types of data are crucial to break 
the degeneracies between the parameters.  So we also consider 
cosmic microwave background (CMB) measurements of the distance to the last 
scattering surface, and a prior on the matter density $\Omega_m$ from 
large scale structure surveys.  The role of CMB complementarity with SN 
measurements was emphasized in \cite{fhlt}. 

\subsection{Method} 

Our parameter set includes ${\cal M}$ -- a ``nuisance'' parameter combining 
the supernova absolute magnitude and the Hubble constant -- and the 
matter density $\om$, present dark energy equation of state $w_0$, 
and time variation $w'$.  We assume a flat universe, which current CMB 
data are consistent with but future measurements will further refine. 
As our fiducial cosmology we adopt $\om=0.28$ and a cosmological constant. 
For the CMB prior we take the WMAP determination of the reduced distance to 
last scattering measured with a 3.3\% relative
precision. 
Similarly, a gaussian prior of 0.03 on $\om$ is used.  The major role of 
the priors is to break degeneracies between parameters rather than 
determine $w'$ per se. 
Neither of these priors is fully descriptive of the complementary probe 
data, but are adequate approximations for the level of precision desired 
for the estimation of $w_0$ and $w'$. 

The time variation $w'$ is taken from the form for the dark energy 
equation of state 
\beq 
w(z)=w_0+w_a(1-a), 
\eeq 
where the scale factor $a=(1+z)^{-1}$, so $w'=w_a/2$ when evaluated 
at $z=1$.  (This value is what we mean by $w'$ for the rest of the paper.) 
This parametrization is very successful in modeling a great 
variety of dark energy (and extended gravity) models 
\cite{linprl,lingrav} and $z=1$ is a 
natural epoch to measure a characteristic time variation.  At higher 
redshifts, in the matter dominated era, dark energy 
is often very slowly varying, while at later 
times many models approach the cosmological constant value of $w=-1$. 
This parametrization allows high redshift data from supernovae and the 
CMB to be combined easily with lower redshift data, since $w(z)$ is 
bounded and well behaved.  Note that the 
parametrization $w(z)=w_0+w_1z$, which is unbounded at high redshifts, 
is unphysical for the CMB data, and a poor approximation for SN data 
at $z>1$. 

For sufficiently large positive $w_a$ the asymptotic value 
of the dark energy equation of state at high redshift $w(z\gg1)=w_0+w_a$ 
will become positive.  This would break matter domination and cause 
problems with, e.g.\ structure formation, and so these models are 
disfavored.  This could be either physically true or an indication 
that the parametrization incorporates the CMB prior poorly.  This 
ambiguity is not a problem if the data 
constrain $w_a\la 1$. We find that in weak data cases 
the Monte Carlo estimations of the parameter uncertainties stray 
into a small region of phase space with $w(z\gg1)>0$, making the estimate 
of the positive (plus) error on $w'$ unreliable.  In those cases we will give 
the parabolic uncertainty (local estimation assuming a quadratic
shape around the minimum) and the 
negative (minus) error 
(see discussion below regarding robustness of the majority 
of the Monte Carlo contour). 
Despite these slight flaws in the parametrization when dealing with weak 
data constraints, the affected regions of parameter space are much smaller 
than in the $w_1$ case, which is more severely pathological and 
runs into trouble even for the SN sample. 

Our fiducial, illustrative supernova sample comprises 40 local SN, placed 
in a bin centered 
at $z=0.05$, and 10 medium redshift SN per bins of redshift width 
0.1 centered at $z=0.15$, 0.25,\dots 0.85.  This gives 120 SN.  We 
supplement this with a high redshift sample of 1 per bin for $z=0.95-1.65$. 
Again, we emphasize this does not represent an actual data set, just a broad 
description of the types of data recently available.  We examine in \S3 
the consequences of increasing the numbers of SN in each of the three 
subsamples. 

Because we expect errors in $w_0$ and $w_a$ to be large, a Fisher method 
of analysis may not guarantee accurate estimations.  We employ both 
Fisher and 
Monte Carlo simulations to obtain our results; Fisher methods 
allow rapid exploration of changes in 
the supernova numbers and systematic errors, providing a guide to 
interesting areas that are then confirmed through Monte Carlo analysis. 

Statistical uncertainties of 0.15 magnitudes, including intrinsic 
dispersion and measurement errors, are placed on each SN and 
then a systematic floor $dm$ is placed on the error in each redshift bin.  
The two types of errors are added in quadrature and the errors are 
assumed uncorrelated from bin to bin.  So for $N$ SN in a bin, 
the total magnitude error per bin is 
\beq 
dm_{\rm tot}=\sqrt{dm^2+0.15^2/N}.
\eeq 
We will see that the systematic uncertainties play a key role.  As a fiducial 
model we take the magnitude systematic to be $dm=A_0+Az$.  Including the 
presence of systematics at low redshift, $A_0\ne0$, is a crucial element 
since this has a strong effect on ${\cal M}$ and, through 
its correlations, on the other parameters.  
This model is clearly a cartoon representation of a very 
complex error propagation problem, but it is tractable and 
should give a not unreasonable estimation \cite{klmm}.  We emphasize that we 
do not claim the systematics model represents any current or planned 
experiment.  No missions were harmed during the writing of this paper. 

Because of 
the excitement of a first estimate of the time variation $w'$, we 
concentrate on this parameter, marginalizing over the others, 
quoting both the negative and parabolic (or positive, where available) 
estimation uncertainties; see the figures for the Monte Carlo contours.  
%Due to the change in sensitivity between negative and positive 
%excursions in $w'$ (as explained in \S\ref{sec.future}), the positive 
%error should lie between the absolute magnitudes of the negative and 
%parabolic. 
For certain cases we present as well the limits that would be 
produced on $w_0$ or a constant $w$ (with $w'$ forced to zero). 

\subsection{Previous data} 

First consider the low and medium redshift SN only. 
Without priors, and taking $dm=0.03+0.07z$, we cannot reasonably 
determine $w'$ simultaneously with the other parameters.  Upon inclusion 
of the CMB and $\om$ priors, we obtain a negative estimation uncertainty 
of $\swp=1.46$ (1.14 parabolic).  
Note that if $A=0.12$ instead of 0.07 then the limit is 1.22 parabolic.  
There are few enough SN that the systematic floor does not have a 
big effect.  So with such data we have not 
restricted $w'$ more tightly than our ``naturalness'' argument of 
Eq.\ (\ref{eq.wdot}).  Data leading to results improving on that ad hoc 
bound would be a welcome advance. 

\subsection{High redshift supernovae} \label{sec.hiz}

Now let's add the sample of high redshift ($z>0.9$) SN.  For 
SN at $z>0.9$ we take $dm=0.1z$.  Without 
external priors $w'$ still cannot be determined; with the priors 
we find $\swp=1.05$ (0.90 parabolic) 
(from now on we always include the priors).  
For the higher systematics case, $dm=0.03+0.12z$ for $z<0.9$ and 
$dm=0.15z$ for $z>0.9$, the parabolic errors jump to 1.01. 

So the high redshift SN (with priors) begin to bring our knowledge 
below the naturalness bound -- a useful step.  But we will see that 
for further improvement we simultaneously need to understand low 
redshift supernovae as well, 
to keep the systematics small, as discussed in \S\ref{sec.future}.  

\subsection{High accuracy supernovae} 

High redshift supernovae, keeping magnitude uncertainties within the 
systematic bounds considered, require observations from space.  Currently 
the Hubble Space Telescope (HST) is the only such data source, and has 
demonstrated an exciting ability to discover and characterize small samples 
of high redshift SN \cite{riess04}. 

Moreover HST has also contributed accurate observations of 
medium redshift SN that have proved excellent for constraining systematic 
uncertainties such as dust extinction \cite{knop}.  We examine the role 
that this type of data plays.  To focus on this last aspect, in this 
subsection we do not include 
the high redshift SN (but do in \S3). 

If the systematics uncertainty is reduced to $dm=0.03+0.04z$ through such 
observations, then the constraint on $\sigma(w')$ (with priors) becomes 
1.08 parabolic.  Such an 
improvement in systematics, without high redshift SN, 
provides roughly 1/4 of the improvement that adding the SN out 
to $z=1.65$, as in \S\ref{sec.hiz}, would.  
Of course this progress is greater -- roughly 2/3 as good an improvement -- 
if the systematics in \S\ref{sec.hiz} were at the higher level considered. 
So high accuracy measurements are a valuable criterion 
in addition to high redshift 
measurements in our quest to understand the nature of dark 
energy. 

\section{Future improvements} \label{sec.future}

Both accuracy and redshift range play a critical role, we just saw. 
HST has enabled strong advances on these two fronts, with the existence 
of the \cite{riess04} and \cite{knop} data, though again we do not view our 
models as representing any real data set.  
If both improvements are combined, using $dm=0.03+0.04z$ for $z<0.9$ 
and $dm=0.1z$ for $z>0.9$, the uncertainty 
$\swp$ improves slightly to 1.0 (0.87 parabolic).   
Understanding of low redshift SN, which functions both to limit 
systematics and essentially calibrate ${\cal M}$ (and hence reduce 
degeneracies with other parameters), is a goal of the Nearby 
Supernova Factory project now underway \cite{snf}.  We simulate this 
forthcoming information by 
taking $dm=0.01+0.06z$ for $z<0.9$.  The numbers then become 
$\swp=0.89$ (0.77 parabolic), demonstrating the important roles of systematics 
control and low redshift SN.  Note there is considerable 
asymmetry between the parabolic and negative (and hence positive) 
uncertainties.  The uncertainty on the present equation of 
state $w_0$ is 0.30.  If one forces $w'=0$, then the 
constant equation of 
state is estimated within 0.12 (but gives very limited physical 
insight because of confusion in interpretation \cite{recon}, since 
we don't know a priori that $w'=0$).  

\begin{figure}[t]
\begin{center}
  \includegraphics[scale = 0.45, trim = 25 10 0 45]{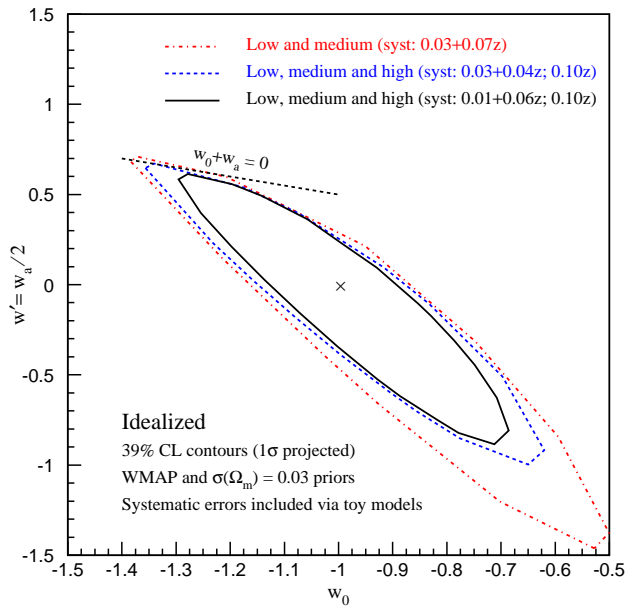}
  \caption{Contours show Monte Carlo estimations of the dark energy 
    equation of state today, $w_0$, and a measure of its time variation, 
    $w'=w_a/2$, for various types of supernova 
    data sets and systematic uncertainties. 
    The contours enclose 39\% of the probability so $1\sigma$ estimation 
    of the parameters is found by projection to the respective axis. 
    Contour points near the $w_0+w_a=w(z\gg1)=0$ boundary line 
    are not robust but the vast majority of the contour is valid. 
  }
  \label{fig.contour1}
\end{center}
\end{figure}

Several of the Monte Carlo contours discussed are plotted in Fig.\ 
\ref{fig.contour1}. 
Asymmetry between the positive and negative 
excursions of the parameters is evident, even excluding from 
consideration the highly 
positive $w'$ region containing a small part of the contour. 
We indicate the cut off 
in the upper left region where we should not believe results as they 
pile up near the $w(z\gg1)\ge0$ boundary. 
With improved data the 
contours will become wholly distinct from the problematic region (the 
best fit should lie elsewhere since our universe did form structure). 

\subsection{Idealistic estimations} \label{sec.ideal}

Can we look forward to rapid, further 
improvement in our knowledge beyond the leap made in the last year, through 
statistical gains as more supernovae are analyzed (other high redshift 
supernovae have already been found)? 

Suppose we increase by a factor 10 the sample at low/medium/high redshift.  
We adopt the model 
$dm=0.01+0.06z$ for $z<0.9$, $dm=0.1z$ for $z>0.9$ from now on. 
The parabolic uncertainties in $w'$ become 0.68/0.65/0.69 or approximately 
10-15\% improvements.  
The systematics floor severely limits any statistical 
improvement.  

So while the most recent HST data has led to an appreciable gain in our 
knowledge, we cannot expect a further dramatic improvement through merely 
gathering more supernovae.  The recent advances have brought us to the 
threshold of the 
systematics limited era of revealing dark energy.  This is an important 
and sobering conclusion.  Let's investigate 
this further. 

\subsection{More realistic estimations} \label{sec.real}

The comprehensive supernova programs over the next several years 
(e.g.\ \cite{essence,snls}) may 
indeed allow a tenfold increase in the low and medium redshift samples, and 
we also consider a more challenging tenfold increase in the $z>0.9$ sample 
in addition. 
Before we quantify the effects of all these together, however, we 
recognize that when we approach higher precision in 
parameter estimation we need to take into account more seriously 
other systematics, which can again inflate the uncertainties.  

For one thing, the heterogeneity of the data samples 
will both increase the dispersion and can lead to biased results. 
Joining together diverse SN data,
taken with different instruments under different conditions can lead 
to a coherent
systematic, basically a calibration offset. This can have a pernicious
influence through biasing the cosmological parameter estimation (see the
discussion in \S4 of \cite{klmm}, for example).  While many samples may
contribute to the tenfold increase in SN data, we assume sufficient
uniformity that we only worry about an offset between local and medium
redshift SN, and medium and high redshift.  One can view these as arising
either from greatly differing telescopes, e.g.\ wide field, small aperture
vs.\ smaller field, larger aperture vs.\ space based, or from different
filter sets, e.g.\ including U, Z, I, J, etc.\ beyond the usual B, V bands.
For a toy model we consider coherent magnitude offsets of 0.02 between 
the low and medium redshift samples and 0.04 between the medium and high 
samples.  Biases from those offsets would amount to a nonnegligible 
$0.5-1\sigma$ shift in the cosmological parameter estimations.  While 
serious, this effect is not included in our quoted $\sigma(w')$ precisions 
in the remainder of this paper. 

A variation is to consider the case where the possibility of such a 
systematic is recognized, and such offsets are incorporated as 
fitting parameters. 
As a toy model we allow new parameters describing these magnitude 
shifts to float subject to gaussian constraints of 0.02 (low-medium 
redshift samples) and 0.04 (medium-high) magnitudes.  In this case 
the cosmology will not be biased from the true model but the parameter 
estimations will suffer increased dispersion.  Allowing for a 10 times 
sample increase for low, medium, and high SN (corresponding to 400 SN 
at $z<0.1$, 100/bin for $z=0.1-0.9$, and 10/bin for $z=0.9-1.7$, for a 
total of 1280 SN) and 
incorporating this calibration systematic yields an
uncertainty of $\sigma(w')=0.58$.  (Now the
contours are small enough to be wholly valid: the positive error
is +0.52, the negative is -0.64, and their mean agrees with the parabolic
error of 0.58). 

An experiment that spectrophotometrically ties the local
and medium redshift SN and creates a homogeneous sample over the entire
range $z=0.1-1.7$, necessarily space based, removes both 
versions of this systematic 
problem and is required for tighter constraints. 

A second major systematic involves the current dominant source of 
systematics uncertainty: 
correction for dust extinction.  This uncertainty may well grow as we 
include SN appearing in higher redshift galaxies.  For example, the 
value of the coefficient in the dust reddening law, $R_V$, may change 
or possess increased scatter, which in turn affects the corrected peak 
magnitude.  (See, for example, \cite{mortsellgoobar}.)  
Here we again neglect the coherent effect which may cause 
bias (see discussion in the conclusion section) and concentrate on the 
incoherent scatter. 

The toy model here takes a magnitude systematic of 0.02 or 0.05, correlated 
within a redshift bin but uncorrelated from bin to bin.  This can be 
viewed as an uncertainty on $R_V$ of 0.2 (or 0.5) with a color excess 
parameter fixed at $E(B-V)=0.1$.  These numbers seem reasonable as, for 
example, \cite{rv} finds that $R_V$ can easily be different from the 
fiducial 3.1 by $\pm0.5$.  The presence of such a systematic degrades 
the estimation of $w'$ for the tenfold increased sample (1280 SN) by 
19\% (63\%) for 0.02 (0.05) mag scatter.  The effect on $w_0$ is even 
more pronounced, at 31\% (106\%). 

Again, a survey that tightly characterizes 
the SN photometry over multiple wavelengths bands from the blue to the 
near infrared, hence space based, strongly constrains the wavelength 
dependence of the reddening, i.e.\ $R_V$, and 
this source of error. 

Treating the calibration and extinction systematics as the two main 
uncertainties, we can address within our toy models what is the most 
``optimistically realistic'' prospect before the next generation of 
experiments 
for determining whether dark energy is dynamical.  We view this 
optimistically as coming from a set of 1280 SN (though as shown this 
is already systematics saturated so more SN will not change the conclusions) 
covering the range $z=0-1.7$ with a calibration systematic as discussed 
above and an extinction systematic at the lower, 0.02 mag value. 
Our baseline prediction then is that $\sigma(w')=0.68$ (as well the 
uncertainty 
on $w_0$ is 0.28 and for a forced constant $w$ is 0.11).  If one were 
less optimistic and used 0.05 mag for the extinction systematic then 
$\sigma(w')=0.90$.  Since Planck CMB data should be available on about the 
same time scale as the tenfold increase in the SN sample, we note 
that using Planck data instead of WMAP brings $\sigma(w')$ to 0.57 (0.79 
with 0.05 mag extinction systematic), though still $\sigma(w_0)=0.27$. 

\begin{figure}[t]
\begin{center}
  \includegraphics[scale = 0.45, trim = 25 10 0 45]{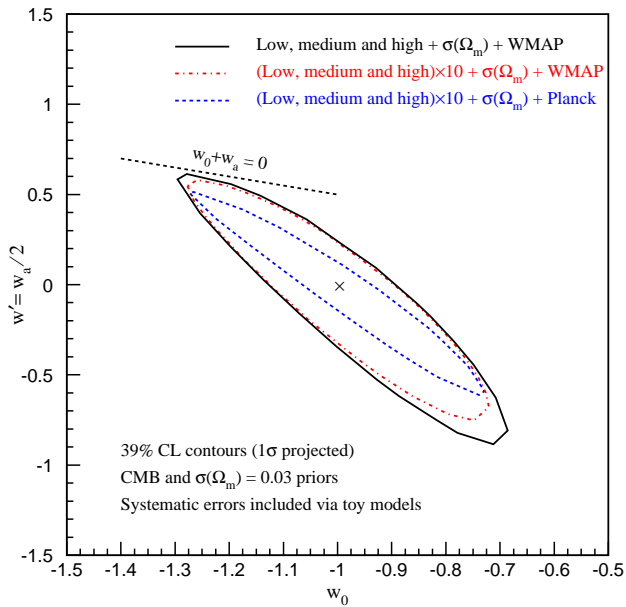}
  \caption{As Fig.\ \ref{fig.contour1}, 
    for various types of data sets 
    that might become feasible through future experiments.  The 
    solid black curve is the same as in Fig.\ \ref{fig.contour1}; 
other curves take 10 times the statistics but include 
two further sources of 
systematics: data sample heterogeneity and extinction correction scatter. 
}
\label{fig.contour2}
\end{center}
\end{figure}

Figure \ref{fig.contour2} illustrates the parameter estimations given 
the vastly enlarged data sets, but also taking into account 
the two systematic effects of heterogeneity of 
data sets and uncertain extinction corrections.  As always, we 
caution these are merely illustrative, toy models.  
The tenfold increase in the $z>0.9$ sample provides little of the 
improvement, because of
the systematics wall -- pointing up the crucial role of systematics
reduction.  
Our most optimistic near term prospect of $\sigma(w')=0.57$ 
represents a possible limit below the
``naturalness bound'' of $1.75\sigma$ (if the data are best
fit by $w'=0$).

To distinguish 
the various dark energy theories common in the literature, however, 
to obtain practical knowledge of the nature of dark energy, 
will require a further generation of SN experiment, with tight systematics 
control over the entire redshift range from 0.1-1.7, say.  In particular, 
we emphasize the vital significance of the improvements that would be 
obtained.  For example, a derived value of $w'=0\pm0.57$ gives 
a less than $2\sigma$ favoring of the cosmological constant below 
the naturalness 
bound -- scarcely sufficient for a problem of such critical implications. 
A next generation space based experiment dedicated to systematics control 
could realistically attain (in combination with Planck) $\sigma(w')<0.15$ 
(cf.\ \cite{omni}). 
This would provide a ``separation power'' of $w'=0$ from $w'=1$ 
at $6.7\sigma$ significance, far more 
robust and worthwhile for such fundamental physics.  
Na{\"\i}vely, in these terms the prospect without a next generation 
corresponds to having a $\sim10$\% ``unconfidence'' 
level in whether dark energy is dynamical, while 
the next generation could reduce this to $10^{-11}$ -- quite a leap forward. 

Finally, we emphasize that this has been only an illustrative look at 
the prospects of exploring the dynamic nature of dark energy.  The 
quantitative results here are dependent on the very simple characterization 
of the SN distribution, priors, and especially systematic uncertainties, 
and should not be taken as more than a broad indication of trends. 

\section{Conclusion} \label{sec.concl}

A great step in our knowledge of dark energy involves the characterization 
of its dynamical properties, i.e.\ the value of the time variation in its 
equation of state.  The first step is whether it is consistent with a 
cosmological constant possessing no variation, i.e.\ limited below 
a ``naturalness'' bound $w'\lesssim1$.  
Measurements of supernovae with the Hubble Space Telescope can achieve 
this reduction, by extending the reach of the magnitude-redshift data to 
higher 
redshifts and by increasing the accuracy of the characterization of 
the supernovae.  Such data should provide limits roughly around 
$\sigma(w')\approx0.8-0.9$. 

Such a step forward brings us near to a systematics wall, however, 
at least at high redshifts, diminishing the prospects for further 
improvements by strengthened 
statistics.  The generation of surveys just beginning, however, will 
provide more data to learn about supernovae, blending valuable 
measurements at low and high redshifts, from both ground and space 
based instruments. By 
identifying and constraining systematic uncertainties they pave the 
way for the next, space based generation to 
obtain tight, physically revelatory bounds on the dark energy dynamics. 
As well, improvements will come in the external priors.  

We caution that systematic uncertainties play a major role in 
translating the observations to knowledge of the nature of dark energy. 
One example is gravitational lensing: the true magnitude of any 
single SN at some redshift is strongly uncertain since gravitational lensing 
by structure in the universe imposes a divergent variance on the 
magnitude; averaging over several SN is required to bring the highly 
nongaussian dispersion down to near the intrinsic SN magnitude 
dispersion (\cite{holzlinder}, also see \S7 of \cite{klmm}, and 
\cite{amanullah}). 

Another caution is apparent from Figure 13 of \cite{knop}.  Systematic 
uncertainties that act to dim the supernovae, such as dust extinction, 
tend to drive the derived equation of state $w(z)$ 
more negative.  If not fully accounted for, the 
data may indicate a more strongly negative $w(z)$, for example a $w'<0$, than 
for the true nature of the dark energy.  (Conversely, overcorrection can 
lead to $w'>0$).  The coherent extinction systematic mentioned in 
\S\ref{sec.real} in fact can bias a true cosmological constant to appear 
as a strongly $w_0<-1$, $w'>0$ (or $w_0>-1$, $w'<0$) model. 
Further investigation of 
a variety of biases is a subject of future work. 

Models with $w'<0$ do exist, such as the linear potential 
model, and can lead to future collapse as discussed in the ``cosmic 
doomsday'' scenario of \cite{kallosh}. 
Whether $w'>0$ or $w'<0$ thus has profound future impact, 
so to learn whether the fate of our universe is collapse or eternal 
expansion first requires stringent control of systematics. 
The dramatic step beyond the $w'$ bounds of the next few years 
will come from space, from missions dedicated to systematics control.  
Quality not quantity is the watchword. 
We emphasize that an improvement from $\la2\sigma$ to $\ga6\sigma$ detection 
of dynamics, or its lack, is a truly dramatic advance, and crucial to 
our understanding. 

In addressing the question of whether dark energy is dynamic, we 
have moved from ignorance to possessing 
the capabilities to limit the more extravagant possibilities, 
and in the third generation begin precision distinction 
of the nature of dark energy.  This will bring us much closer to 
knowledge of fundamental physics and the fate of the universe, whether 
deSitter expansion under 
the cosmological constant, superacceleration and a Big Rip under phantom 
energy (but see \cite{lingrav} for an opposing conjecture about fate 
under superacceleration), or collapse under ``cosmic doomsday'' 
scenarios. 

\section*{Acknowledgments} 

We thank Greg Aldering, Alex Kim, Saul Perlmutter, Adam Riess, Lou 
Strolger, and David Weinberg for helpful conversations. This work has 
been supported 
in part by the Director, Office of Science, Department of Energy under 
grant DE-AC03-76SF00098.  RM is partially supported by the National 
Science Foundation under agreement PHY-0355084.  EL acknowledges the 
Aspen Center for Physics for hospitality during completion of this work.

\end{document}